\documentstyle{mn}
\title {
General Relativistic Collapse of Neutron Stars to Strange Stars: A
Mechanism for Gamma Ray Bursts}
\author[A. Mitra]      {Abhas Mitra\\
Theoretical Physics Division, Bhabha Atomic Research Centre,\\
Mumbai-400085, India\\ E-mail: amitra@apsara.barc.ernet.in}
\date{
}
\pagerange{\pageref{firstpage}--\pageref{lastpage}}
\pubyear{2000}
\begin{document}
\maketitle
\label{firstpage}
\begin{abstract}
It is known that  Neutron Stars may be converted into more compact Strange
Stars (SS) on capture/formation of a ``seed'' of strange matter.  It is
also known that the binding energy of the nascent hot SS is likely to be
radiated as $\nu -\bar\nu$s so that an electromagnetic pair fireball (FB)
may be created by neutrino annihilation.  But we show here that,   a
General Relativistic treatment of the problem may lead to a FB energy
($Q_{FB}\sim ~10^{53}$ erg) which could be  higher by as much as {\em four orders} than
what was estimated previously.  Further since the baryonic mass of the
thin crust of a strange star is negligible, this FB will generate an
extremely relativistic blast wave.  Thus this process may be one of the
viable routes for the genesis of hitherto unexplained cosmic Gamma Ray
Bursts.
\end{abstract}
\begin{keywords}
 Gamma rays: bursts -- Gamma rays: theory Stars: neutron  -- Relativity --
Gravitation
\end{keywords}
\section{Introduction}
It is now clear that a large number of Gamma Ray Bursts (GRBs) involve
emission of $\gamma$-rays as large as  $Q_\gamma \sim 10^{52} -10^{54}$
erg under condition of isotropy (Kulkarni et al. 1999). However, many GRB afterglow light
curves  decline more rapidly than what is expected in simple isotropic
fireball models, and it is plausible that such afterglows are actually
beamed (Kulkarni et al. 1999, Huang et al. 2000). In such cases, the total (electromagnetic) energy budget
could be substantially reduced (often by a factor $\sim 100$).
Nonetheless, many afterglows do not fade with such unusual rapidity and
the decay of the light curve may be smoothly fitted by a single power on
the time scale of months. Such GRBs are likely to be more or less
spherical events, and thus,  for GRB 971214, we may indeed have $Q_\gamma
\approx 3\times 10^{53}$ erg (Dal Fiume et al 2000). There have been many attempts to
explain the energy budget and origin of powerful GRBs with especial
emphasis on the beamed  (Paczynski 1998, Popham, Woosley \& Frier 1999,
Ruffert et al. 1997).  In the following, we put forward an alternative
model to explain  the origin of GRBs with a value of  $Q_\gamma \sim
10^{53}$ erg for which the problem of baryonic contamination is minimal
because we are considering the possibility of conversion of an {\em
isolated} or mass accreting massive neutron star (NS) into a Stange Star
(SS). It might appear at first sight that the model  is not  new at all
because starting with Alcock, Farhi \& Olinto (1986), several other
authors have considered this process as a probable mechanism of GRBs (Ma
\& Xie 1996, Cheng \& Dai 1996).  For example, Ma \& Xie (1996) studied
the phase transition of a neutron  star into a ``hybrid'' star. Cheng \&
Dai (1996) also pointed out that a NS in a low mass X-ray binary, after
accreting sufficient mass may undergo a similar phase transition and
power a weak GRB event. While Alcock, Farhi \& Olino attempted to explain
an electromagnetic FB of $Q_{FB} \sim 10^{45-46}$ erg, the other authors,
crudely estimated that the value of $Q_{FB}\sim 10^{49}$ erg.  Clearly
such low values of $Q_{FB}$ are highly insufficient to explain the
energetics of presently observed GRBs. Thus while the basic concept
involved in this paper is not new ($NS \rightarrow SS$), this general
relativistic model as such is new because it may explain four order higher
values of $Q_{FB}$.

Since this paper has similarity in idea with that of Cheng \& Dai (1996),
we shall briefly recall the main result of the latter paper. Cheng \& Dai
adopted a different approach and estimated that the new born strange star has
a thermal energy $E_{\rm th} \sim 5 \times 10^{52}$ erg. The star cools by
emitting this thermal energy in the form of $\nu, \bar\nu$ pairs. While
the neutrinos traverse out from the core, they impart electromagnetic energy
inside the star by processes like $n + \nu_e \rightarrow p + e^-$ and $p+
\bar \nu_e \rightarrow n+ e^+$. Cheng and Dai estimated the optical
thickness of the above processes to be $\tau \sim 4.5 \times 10^{-2}$ and
concluded that the electomagnetic energy deposited {\em within the body of
the star} to be $E_2 \sim \tau \times E_{\rm th} \sim 2\times 10^{52}$
erg , which is obviously incorrect for the given value of $E_{\rm th}$ and $\tau$.  Thus, as one can easily verify, Cheng \& Dai here overestimated
$E_2$ by a factor of 10 and its value works out to be $E_2 \sim 2 \times
10^{51}$ erg. Moreover, Cheng \& Dai overlooked here fact that pairs
generated {\em within the body} of the star can not come out because the
star is exceedingly optically thick (electromagnetic optical thickness is
$\sim 2.5$ g cm$^{-2}$. Thus this electromagnetic energy gets absorbed
within the star and reemitted primarily as neutrinos. So the total
neutrino flux coming out of the star is $\sim E_{\rm th} \sim 5 \times
10^{52}$ erg. Now Cheng \& Dai esitimated that total energy coming out
from the star in the form of pairs, due to the $\nu +\bar\nu \rightarrow
e^+ + e^-$ process, is $E_1 \sim 10^{49}$ erg. Thus, the total {\em external}
 pair flux flux energy, in the model of Cheng \& Dai should be $E_0
\approx E_1 \sim 10^{49}$ erg and not $E_0 = E_1 + E_2 \sim 5 \times
10^{52}$ erg as erroneously concluded by them.

The general relativistic treatment presented below would show that,
actually, the pair fireball (FB) enegy can be much more.

\section{The Model}

We consider an initial scenario where a NS, either isolated or accreting
mass in a low mass X-ray binary, is of low compactness. This is possible
for a class of low density stiff NS equation of states.  For instance, the
radius of a $1.4 M_\odot$ NS with a (low) fiducial density parameter
$\rho_0$ could be (Kalogera \& Baym 1996).
\begin{equation}
R=21.2 {\rm km} \left({\rho_0\over 10^{14} {\rm g cm}^{-3}}\right)^{-0.35}
\end{equation}
Here $\rho_0$ is not to be confused with the central density $\rho_c$.
Such low density NSs have a binding energy (BE) insignificant compared to
their baryonic mass $E_B^i \sim 0.1 M_\odot c^2$. Here
$M_\odot$ is solar mass and $c$ is the speed of light.
Following Olinto (1987) we consider  the
possibility  that the NS is hit by a primordial strange matter clump or
seed in the
form a cosmic ray. Once the NS gravitationally captures such a ``seed'',
the neutrons surrounding the seed will be absorbed by it and be deconfined
to be strange matter (Alcock \& Olinto 1988, Cheng, Dai \& Lu 1998). 
Thus the strange matter seed can become bigger and
bigger until the whole star is converted into a strange star. And this
process may be complete extremely fast on a timescale of $10^{-7}$ s
(Alcock \& Olinto 1988, Farhi \& Jaffe 1984)
This process may  occur for an accreting NS once it starts collapsing
after its mass crosses a certain
maximum value and the central density increases sufficiently. This might
also happen for the rapidly spinning supramassive NS scenario (Vietri \&
Stella 1998) where the NS
starts collapsing after loss of sufficient angular momentum.

The maximum values of masses and  radii of strange stars may be
represented by (Witten 1984,  Cheng, Dai \& Lu 1998)
\begin{equation}
M= 2.0 ~M_\odot \left({B_0\over B}\right)^{1/2}
\end{equation}
and
\begin{equation}
R= 11.1 \left({B_0\over B}\right)^{1/2}~{\rm km}
\end{equation}
where the ``bag constant'' is $B$ and $B_0= 57$ MeV/fm$^3$
Therefore  the surface
gravitational redshift of a {\em static}  spherical strange star
is given by (Shapiro \& Teukolsky 1983) :
\begin{equation}
z =   \left (1 - {2G M(R)\over Rc^2}\right)^{-1/2} -1 \approx 0.5
\end{equation}
Here  $r$ is the invariant circumference radius, $c$ is the speed of
light, and $M$ is the gravitational mass enclosed within $r=r$
\begin{equation}
M(r) = \int_0^r \rho dV =\int_0^r dM
\end{equation}
where $\rho$ is the total mass-energy density, $dV=4\pi r^2 dr$ is
coordinate volume element, and the symbol $dM$ is self-explanatory.

The self-gravitational energy of a static relativistic star is given by 
(Shapiro \& Teukolsky 1983)

\begin{equation}
E_g= \int \rho  dV \left\{1- \left[1- {2M(r)\over r}\right]^{-1/2}\right\}
\end{equation}
 Then recalling the definition of $z$ from Eq. (4), we may write
\begin{equation}
 E_g= -\int z(r) dM \approx \alpha z M \approx -z M
\end{equation}
where $\alpha \sim 1$ is a model dependent parameter and studying the
models of relativistic polytropes, we
have numerically verified that for $z<0.6$, indeed $\alpha \ge 1$.

The binding energy of a cold star is given by $E_B \approx (1/ 2)
\mid E_g \mid$. If the NS undergoes a phase transition to become a more
compact SS, the total energy to be liberated would be $E_B (SS) - E_B
(NS)$. But here we are considering {\em only those situations where the
original NS is not very compact} (Kalogera \& Baym 1996) with a canonical value of $z_i \le
0.15$, and thus, for the sake of analytical simplicity we neglect the
initial B.E. of the NS in comparison to the much larger B.E. of the SS,
and to compensate for the initial B.E. we set here $\alpha=1.0$ although,
actually it is marginally higher. Since  occurrence of
powerful GRBs having a frequency $\sim
10^{-6}-10^{-7}$/galaxy/yr (under conditions of isotropy) is 
an extremely rare event  
compared to the same for supernova events, it may be justified
to consider only favourable initial condtions.  Further during the NS-SS
conversion, an additional energy of $\sim 30$ MeV/nucleon is liberated due
to quark deconfinement (Alcock \& Olinto 1988, Fari \& Jaffe 1984). For the sake of analytical simplicity,
we club all such liberated energy into a single expression $Q_\nu \approx
E_B \approx {z M \over 2}$.  In the case of NS formation, the nascent hot
NS radiates most of its BE in the form of $\nu-\bar\nu$ on a time scale of
$\sim 10$ s (Shapiro \& Teukolsky 1983) and we take this as a fiducial scale for the present problem.

So, given this value of $z=0.5$ the maximum value of
$Q_\nu
\approx 1.0    \times 10^{54} M_2 $ erg where of $M= M_2
 2 M_\odot$.
The value of $Q_\nu$ measured near the compact object will be higher by a
factor $(1+z)$: $Q_\nu' =  z (1+z) M/2$.
  And the locally measured duration
of the burst would be $t_\nu'=(1+z)^{-1} t_\nu$.  Therefore, the mean
(local) $\nu -\bar \nu$ luminosity    will be
\begin{eqnarray}
L_\nu' &=& {Q_\nu'\over t_\nu'}= {z (1+z)^2 M \over 2 t_\nu}
\approx \nonumber\\
&  & 2\times 10^{53} z (1+z)^2 M_2~ t_{10}^{-1} ~erg/s
\end{eqnarray}
where $t_\nu =t_{10} 10 s$. As to the cooling of strange stars, there is an
important difference with respect to a NS: while a NS cools via emission
of all 3 flavours of $\nu,\bar\nu$ of approximately equal proportion, a SS
cools primarily {\em via the emission of electronic} $\nu_e, \bar\nu_e$:
\begin{equation}
u+e^-\rightarrow d+\nu_e,~ u+e^- \rightarrow s+\nu_e, ~d\rightarrow
u+e^-+\bar\nu_e, ~s\rightarrow u+e^-+\bar\nu_e
\end{equation}
 Thus
there will primarily be {\em only two specieses of neutrino} each contributing
to a luminosity
 $L_i' =(1/2) L_\nu'$.  In contrast normal baryonic matter (like a NS) cools via emission of six specises of neutrinos. Consequently, emissivity of electromagnetic pairs is much higher for a SS cooling in comparison to a NS cooling, and we feel that, this  point has not been noted earlier. By assuming the radius of the
neutrinosphere to be $R_\nu \approx R$, the value of effective local
neutrino temperature $T'$ is
obtained from the condition
\begin{equation}
L_\nu' = {7\over 8} 4\pi R^2 \sigma T'^4
\end{equation}
where $\sigma$ is the Stefan-Boltzman constant. By equating Eqs. (8) and
(10), we find
\begin{eqnarray}
T'  &= &\left( {4  z (1+z)^2 M c^2\over 7 \pi \sigma R^2
t_\nu}\right)^{1/4} 
 \sim 18 MeV \nonumber\\
 & & z^{0.25} (1+z)^{0.5}
M_2^{0.25} ~R_6^{-0.5} ~t_{10}^{-0.25}
\end{eqnarray}
where $R= R_6 10^6$.  For a Fermi-Dirac distribution, under the crude
assumption of zero $\nu$-chemical potential, the mean (local) energy of
the neutrinos is $E_\nu' \approx 3.15  T' \sim 57$ MeV (for $z=.5$).  The
various neutrinos will collide with their respective antiparticles to
produce electromagnetic pairs by the $\nu +\bar\nu \rightarrow e^+ + e^-$
process. The rate of energy generation by pair production per unit volume
per unit time,  at a distance $r$ from the center of the star, is given by
(Goodman, Dar \& Nussinov 1987)
\begin{equation}
{\dot q}_\pm (r) = \sum_i {K_{\nu i} G_F^2 E_\nu' L_i'^2(r) \over 12
\pi^2 c R_\nu^4} \varphi(r)
\end{equation}
Here, $L_i'(r)\sim r^{-2}$ is the $\nu$-flux density of a given species
above the $\nu$-sphere, $G_F^2= 5.29 \times 10^{-44}~cm^2~ MeV^{-2}$ is
the universal Fermi weak coupling constant squared, $K_{\nu i} =2.34$ for
electron neutrinos and has a value of 0.503 for muon and tau neutrinos.
Here the geometrical factor $\varphi(r)$ is
\begin{equation}
\varphi(r) = (1-x)^4 (x^2 +4x +5); \qquad x= [1- (R_\nu/r)^2]^{1/2}
\end{equation}
But since in the present case, most of the neutrinos are of the electronic
nature, there is {\em substantial enhancement of the efficiency}:
\begin{equation}
{\dot q}_\pm (r) =  {K_{\nu,e } G_F^2 E_\nu' L_\nu'^2(r) \over 48
\pi^2 c R_\nu^4} \varphi(r)
\end{equation}
Now, a simple numerical integration yields
the local value of pair luminosity produced above the neutrinosphere :
\begin{eqnarray}
L_\pm' &=& \int_R^\infty {\dot q}_\pm 4 \pi r^2 dr \approx
\frac{K_{\nu,e } {{G_F}^2}  E'_{\nu} {L_{\nu}^{'}}^2}{27 \pi c
R_\nu} 
\approx 1.2\times  10^{52} \nonumber\\
& &~z^{2.25}~(1+z)^{4.5}~
M_2^{2.25}~t_{10}^{-2.25}~R_6^{-2} erg/s
\end{eqnarray}
This estimate is obtained by assuming rectilinear propagation of neutrinos
near the SS. Actually, in the strong gravitational field near the SS
surface the {\em neutrino orbits will be curved} with significant higher
effective interaction cross-section. Since, most of the interactions take
place near the $\nu$-sphere, for a modest range of $z$, we may tentatively
try to incorporate this nonlinear effect by inserting a $(1+z)^2$ factor
in the above expression. On the other hand, the value of this
electromagnetic luminosity measured by a distant observer will be smaller
by a factor of $(1+z)^{-2}$, so that eventually, $L_\pm =L'_\pm$ of
Eq.(15). And the total energy of the electromagnetic FB at $\infty$ is
\begin{eqnarray}
Q_{FB} & =& t_\nu ~L_{\pm}\approx 1.2\times 10^{53}\nonumber\\
& &z^{2.25}~(1+z)^{4.5}~ M_2^{2.25} ~t_{10}^{-1.25}~R_6^{-2}~erg
\end{eqnarray}
For $z= 0.5$, $M_2
\approx 1$,  $R_6 \approx 1.1$, and $t =10 s$, we obtain a highest value
of $Q_{FB} \approx 1.5\times 10^{53}$ erg.  The {\em efficiency} for
conversion of $Q_\nu$ into $Q_{FB}$ in this case is $\epsilon_\pm = Q_{FB}
/ Q_\nu
\approx 15\%$.  It is known that NSs have an upper limit of mass $\approx
3.0 M_\odot$ and with the inclusion of rotation the maximum mass could be
as high as $3.7 M_\odot$ (Shapiro \& Teukolsky 1983). Similarly for a bag constant $B=B_0= 57$
MeV/fm$^3$, the maximum mass of a rotating SS is 2.4 $M_\odot$ (Colpi \&
Miller 1992). At this
stage the actual value of $B_0$ is uncertain and thus a value of $M \ge 3
M_\odot$ can not be ruled out for a SS. Then Eq. (16) may yield a maximum
value of $Q_{FB} \approx 3.7\times 10^{53}$ erg, as may be required for GRB 971214.

\section{Discussion}
In some ways, our model in the General Relativistic extension of the
previous work by Cheng \& Dai (1996), and it is free from the ``baryon
pollution problem'' in same way.
The mass of the baryonic crust of a canonical SS is only $\Delta M=
2\times 10^{-5} M_\odot$ (Cheng, Dai \& Lu 1998, Alcock \& Olinto 1984).
For the massive SS, even if the value of $\Delta M$ is considerably
higher, we would always have the degree of baryonic pollution  $\eta=
Q_{FB} /\Delta M > 10^2$, and the genesis of a GRB may be understood
(Meszaros \& Rees 1992).

For quick comprehension by the readers, let us explain here, why our
relativistic model 
is able to generate a FB energy as high as four orders of magnitude than
previous non-relativistic estimates. In a non-relativistic model, one
usually considers a NS with mass of 1 $M_\odot$, so that $M_2= 0.5$. The
radius of such a NS is taken as 10 Km so that $R_6=1$. For such a NS $z
\approx 0.15$. We call this Case I. On the other hand, for the present
relativistic model (Case II) involving a heavy NS with mass 2 $M_\odot$, we have
$M_2 =1$. The radius of such a NS is 11 Km which is practically the same
as in Case I, so that for qualitative understanding, we neglect any
variation with respect to $R$ in the two cases. As mentioned before, in
this latter case $z\approx 0.5$.

Then for a a burst of
duration of 10 s, the essential relativistic fators in Eq. (16) are

\begin{equation}
Q_{FB} \sim z^{2.25} (1+z)^{4.5} M_2^{2.25}
\end{equation}

For case I, the numerical value of this factor is $Q_{FB} \sim 0.005$. And
for Case II, the same numerical factor is $Q_{FB} \sim 1.3$. Thus,
ignoring higher pair emissivity for the moment, the ratio of relativistic
factors in the two cases is 260. The previous authors ignored the fact
that for SS cooling, pair emissivity is much higher compared to the case
of NS cooling.  For the latter case (Case I), for a total neutrino luminosity of
$L_\nu'$, the luminosity in each species is $L_i' = (1/6) L_\nu'$. In
contrast for
the former case (Case II), we have $L_i' = (1/2) L_\nu'$. The pair
emissivity, as per Eq. (12) is $\propto L_i'^2$. The same is also $\propto
E_\nu' \propto T' \propto L_i'^{0.25}$. Then finally, pair emissivity
$\propto L_i'^{2.25}$. Since, for Case II, $L_i'$ is higher of 3, one
would obtain
a pair emissivity higher by a factor of $\approx 12$ here if the value of
$K_{\nu i}$ were same for all the flavours of neutrinos. But, as mentioned
before, while $K_{\nu i} = 0.503$ for $\mu$ and $\tau$ neutrinos, its
value is 2.34 for electronic neutrinos. This fact enhances the pair
emissivity approximately by a factor of 4 for Case II. Considering all the
three effects, FB energy becomes higher by a net factor of $260\times
12\times 4 \approx 10^4$ in Case II!

It might be possible that upon capturing a sufficiently massive
strangelet, not only a NS, but an ordinary star like the Sun can also get
converted into a SS (Alcock \& Olinto 1988). Here the burning process into
strange matter arises via a series of proton capture reactions apart from
the usual neutron capture and deconfinement mode (Alcock \& Olinto 1988).
If the NS$ \rightarrow$ SS process is complete within $10^{-7}$ s, it is
conceivable that a normal star $\rightarrow$ SS process is completed well
within 1 s. And if a massive star gets converted into a maasive SS in this
way, we would be able to explain origin of GRBs having energy even much
higher than $10^{54}$ erg.  However, the resultant massive SS must be
unstable and is likely to proceed for further collapse.

In the case of a supernova, even in an initial spherical model there could
be symmetry breaking Rayleigh-Taylor and Kelvin-Helmoltz instabilities
because of the  severe density gradients in the presence of extreme and
varable gravity (Burrows 2000).  There could be additional asyymetry if
the  SS is magnetized  and spins fast, and thus it is plausible that the
resultant GRB will be  beamed. Finally, the recent discovery of a
quark-gluon plasma state in CERN provided as an additional motivation
behind this work (Abott 2000).
\section*{Acknowledgements} The author thanks the anonymous referee whose suggestions led to the improvement of the scientific clarity of this paper.


\begin{thebibliography}{}
\bibitem{1}  Abbott,  A., 2000,  Nature, {\bf 403} 581. This is a news
report.


\bibitem{2}  Alcock,  C.,  Farhi.,  E., \&  Olinto,  A.,
1986, ApJ {\bf 310}, 261


\bibitem{3}  Alcock , C. \&  Olinto, A., 1988, l Ann. Rev. Nucl. Part. Sci.
, {\bf
38}, 161

\bibitem{4}  Burrows,  A., 2000,  Nature, {\bf 403}, 727

\bibitem{5}  Cheng, K.S. \& 
 Dai,  Z.G., 1996,  Phys. Rev. Lett., {\bf 77}, 1210

 \bibitem{6}    Cheng,  K.S. ,   Dai,  Z.G. \&  Lu,  T., 1998, 
  Int. J. Mod. Phys.
D., {\bf 7}, 139   

\bibitem{7}   Colpi, M.  \&  Miller,  J.C., 1992,  ApJ, {\bf 388}, 513 

\bibitem{8}  Dal Fiume , D. et al., 2000, l AA, (in press) (astro-ph/0002229)

\bibitem{9}   Farhi, E. \&  Jaffe,  R., 1984,  Phys. Rev. D, {\bf 30}, 2379 


\bibitem{10}  Goodman,  J.,   Dar,  A. \&   Nussinov,  S. 1987,  ApJ.
 {\bf 314},
  L7 

\bibitem{11}  Huang, Y.F.,   Dai,  Z.G.,  Wei,  D.M.  \&  Lu,  T., 2000,
 ApJ (submitted)  (astro-ph/0002433)

\bibitem{12}  Kalogera, V. \&  Baym,  G., 1996, ApJ, {\bf 470}, L61 

\bibitem{13}  Kulkarni, S.R.  et al., 1999, Nature, {\bf 398}, 389 

\bibitem{14} Ma, F. \&  Xie,  B., 1996,  ApJ, {\bf 462}, L63 

\bibitem{15}  Meszaros , P. \&  Rees,  M.J., 1992,  ApJ  {\bf 397}, 570 

\bibitem{16}  Olinto, A., 1987, Phys. Lett. B {\bf 192}, 71 


\bibitem{17}   Paczynski, B., 1998,  ApJ, {\bf 494}, L45 

\bibitem{18}   Popham,  R.,
 Woosley,  S.E., \&   Fryer, C., 1998, ApJ, {\bf 518}, 356   (astro-ph/9806299)

\bibitem{19} Ruffert,  M,   Janka,  H,  Takahashi,  K.\&
 Schafer, G., 1997, A\&A, {\bf 319}, 122 

 












\bibitem{20}  Vietri, M. \&  Stella,  L., 1998,  ApJ, {\bf 507}, L45



\bibitem{21}  Witten, E., 1984,  Phys. Rev. D, {\bf 30}, 272 

\bibitem{22}   Shapiro, S.  \&   Teukolsky,  S.A., 1983, 
 Black Holes, White Dwarfs, and
Neutron Stars: The Physics of Compact Objects, (Wiley, New York)





























\end{thebibliography}
\end{document}